\documentclass[11pt,iop]{emulateapj}
\usepackage{ulem}
\usepackage{apjfonts}

\shorttitle{The bulk Lorentz factors of GRBs}
\shortauthors{Tang et al.}

\begin{document}

\title{Measuring the bulk Lorentz factors of gamma-ray bursts with \textsl{Fermi}}

\author{Qing-Wen Tang\altaffilmark{1,2}, Fang-Kun Peng\altaffilmark{1,2}, Xiang-Yu Wang\altaffilmark{1,2}, Pak-Hin Thomas Tam\altaffilmark{3}}
\affil{$^1$ School of Astronomy and Space Science, Nanjing University, Nanjing 210093, China; xywang@nju.edu.cn  \\
$^2$ Key laboratory of Modern Astronomy and Astrophysics (Nanjing University), Ministry of Education, Nanjing 210093, China \\
$^3$ Institute of Astronomy and Space Science, Sun Yat-Sen University, Guangzhou 510275, China; tanbxuan@mail.sysu.edu.cn \\
}

\begin{abstract}
Gamma-ray bursts (GRBs) are powered by ultra-relativistic jets.
Usually a minimum value of the  Lorentz factor of the relativistic
bulk motion is obtained based on the argument that the observed
high energy photons ($\gg {\rm MeV}$) can escape without suffering
from absorption due to pair production. The exact value, rather
than a lower limit, of the Lorentz factor can be obtained  if the
spectral cutoff due to such absorption is detected. With the good
spectral coverage of  the Large Area Telescope (LAT) on
\textsl{Fermi}, measurements of such cutoff become possible, and
two cases (GRB~090926A and GRB~100724B) have been reported to have
high-energy cutoffs or breaks. We systematically search for such
high energy spectral cutoffs/breaks from the LAT and the Gamma-ray
burst monitor (GBM)  observations of the prompt emission of GRBs
detected since August 2011. Six more GRBs are found to have
cutoff-like spectral feature at energies of $\sim10-500$~MeV.
Assuming that these cutoffs are caused by pair-production
absorption within the source, the bulk Lorentz factors of these
GRBs are obtained. We further find that the Lorentz factors are
correlated with the isotropic gamma-ray luminosity of the bursts,
indicating that more powerful GRB jets move faster.

\end{abstract}


\section{Introduction}

Gamma-Ray Bursts (GRBs) are the most energetic transient phenomena
in the universe. The initial brief and intense gamma-ray flash,
the so-called prompt emission, is thought to be produced in an
ultra-relativistic outflow, as argued by the fact that  high
energy photons ($\gg {\rm MeV}$) escape out of the source without
suffering from absorption due to pair production
($\gamma\gamma \leftrightarrow e^+e^-$) (e.g,
~\citet{1991ApJ...373..277K,1993A&AS...97...59F,1995ApJ...453..583W,1997ApJ...491..663B}).
Requiring that the absorption optical depth
$\tau_{\gamma\gamma}\la1$ for high energy photons, one can deduce
a lower limit on  the bulk Lorentz factor ($\Gamma$) of the
emitting region, which is usually $\ga100$
~\citep{2001ApJ...555..540L}.

The absorption should cause a spectral cutoff or break in the
highest energy end, which is expected to be seen within fireball shock model if the energy
coverage and sensitivity of the detector is sufficiently good.
With the greatly increased  spectral coverage of  the Large Area
Telescope (LAT) on \textsl{Fermi},  search for such high energy
spectral cutoff/break becomes possible. A spectral break around
0.4 GeV is detected for the first time in  GRB
090926A~\citep{2011ApJ...729..114A}. In the first \textsl{Fermi}/LAT GRB
catalog~\citep{2013ApJS..209...11A}, which summarized the spectral
analysis of all LAT-detected GRBs up to July 2011, one more GRB
(i.e. GRB 100724B) is reported to have  a spectral cutoff at the
highest energy end. In this paper, we perform a thorough analysis of GRBs
detected by \textsl{Fermi}-LAT between August 1, 2011 and October 30, 2014
to search for cutoff-like spectral features. In \S 2, we present
the sample selection (\S 2.1), data reduction (\S 2.2) and the results (\S 2.3). We find six out of twenty-eight GRBs showing cutoff-like features, and the rest bursts can be
adequately modeled by the Band function. In \S 3,  assuming the
cutoffs are caused by the pair-production absorption in emission region, the bulk
Lorentz factors ($\Gamma $) are obtained (\S 3.1). With the total eight GRBs having
measurements of the Lorentz factors, we further test the $\Gamma-L_{\rm \gamma, iso}$ and $\Gamma-E_{\rm \gamma, iso}$ correlations (\S 3.2). Then we give a summary (\S 3.3). Throughout this paper, we adopt a Hubble constant $ H_{0}=71 {\rm km s}^{-1} {\rm
Mpc}^{-1}$,$\Omega_{\rm M}=0.27$ and $\Omega_{\Lambda}=0.73$.

\section{Data analysis and results}
\subsection{The burst sample}
Since the launch of \textsl{Fermi} satellite, $\sim 250$ GRBs per
year are detected with GBM. When sources are bright enough, the
spacecraft will slew to the   location of the burst and performs a
pointed observation autonomously. So the prompt emission of some
GRBs were simultaneously observed with LAT. We search for such
GRBs to make joint spectral analysis. A total of 49  GRBs are
detected by the $Fermi$/LAT between August 1, 2011 and October 31,
2014, as listed in the LAT Burst Online
Catalog~\footnote{\url{http://fermi.gsfc.nasa.gov/ssc/observations/types/grbs/lat$\_$grbs/}}.
Focusing on the prompt phase, 28 of them were reported by the
\textsl{Fermi}/LAT collaboration (through GCN circulars) to have
LAT detection (>100~MeV) during the main gamma-ray emission phase,
i.e., the time interval of GBM data analysis. Table 1 shows the
information of these 28 GRBs, including the position derived from
the LAT photons, the burst time interval used by the GBM team for
spectral analysis (which is also the time interval used in our
joint GBM/LAT analysis) and the LAT boresight angle at the GBM
trigger time.

\subsection{Data analysis}
\subsubsection{Data preparation and event selection}
{We extract both LAT and GBM data from the FSSC (Fermi Science
Support Center). During the spectral analysis, the Time-Tagged
Events (TTE), as well as CTIME, data files from two or three NaI
detectors and one BGO detector were used. For 15 out of the 28
GRBs, publicly LAT Low-Energy (LLE) data are also available, as
shown in Table 1. We then performed joint spectral analysis
including LLE data for these 15 bursts. For each NaI detector,
channels below 8~keV or above 1000~keV are ignored. For BGO, we do
not include channels below 250 keV or above 40~MeV.} The time
interval for spectral analysis is GBM $T_{90}$, which contains the
emission from $5\%$ of its total fluence to $95\%$(tabel 1)
~\citep{2012ApJS..199...18P,2012ApJS..199...19G,2014ApJS..211...13V,2014ApJS..211...12G}.
The joint spectral fitting of GBM/LAT data is performed with RMFIT
version 4.3.2., which judge the goodness of fit using the Castor
Statistic(CSTAT) to handle correctly the small number of events at
the high energy.

\subsubsection{Background estimation and spectrum extraction}
1) GBM data.  For each detector, two off-pulse time intervals (one
before and one after the GRB prompt pulses) are selected, and then
we fit them with polynomial functions in RMFIT. The order of the
fitting polynomial was chosen as one in the beginning, incremented
by one each time, until we get a reduced $\chi^2 \simeq 1$, with a
maximum of four. In order to minimize the statistical and
systematic uncertainties (and hence ensure a reliable background
estimate), the off-pulse time intervals must be close to the burst
interval, have a long enough duration, and do not contain bumps or
other structures in the light curve. After each fit, we check
visually that the residual are consistent with the statistical
fluctuations. If not, we repeat the procedure by changing the
choice for the off-pulse intervals. {The CTIME event data are
employed to obtain background models.}

2) LLE data. { The LLE data products are delivered by the LAT team
to the FSSC and to the public which provides large effective area
at low energies of the LAT detector, joining the LAT and GBM
energy ranges. We include the LLE events from 30~MeV to 130~MeV in
the fit, similar to \citet{2012ApJ...757L..31A}.} We estimate its
background using the same procedure as GBM data but with the
publicly LLE spectrum products which can be used in RMFIT and
XSPEC.

3) LAT data (>100~MeV). We derive the LAT spectrum files and
response files using the \textsl{Fermi} Science Tools package, version
v9r32p5~\footnote{available at the Fermi Science Support Center
(FSSC), \url{http://fermi.gsfc.nasa.gov/ssc/}}. We select
transient-class events from LAT observations, and instrument
response functions (IRFs) P7REP$\_$TRANSIENT$\_$V15 are used. We
excluded the events with zenith angles $>$100$^\circ$ in order to
avoid a significant contribution of Earth-limb gamma-rays using
the tool $gtselect$. All events in a region of interest (ROI) of
12$^\circ$ around the positions(see Table. 1) of burst are used.
LAT spectra are divided into 20 logarithmically spaced energy bins
from 100~MeV to 10~GeV. The spectrum and the response matrix of
each GRB are derived using $gtbin$ and $gtrspgen$. Background of
LAT spectra are calculated by the BKGE script with a constant ROI
by adding a command of $"ROI\_Calculate=0"$~\citep{2013APh....48...61V}.

\subsubsection{Spectral models}
GRB time-integrated spectra are usually fitted with a smoothed
broken power law function, the so-called "Band
function"~\citep{1993ApJ...413..281B}.  {Possible superposition of
a thermal component on the nonthermal  spectrum was claimed in
\textsl{BATSE} and \textsl{Fermi}
GRBs\citep{2011ApJ...727L..33G,2013ApJ...770...32G,2015arXiv150107028G,
2014Sci...343...51P,2009ApJ...702.1211R,2005ApJ...625L..95R,2010ApJ...709L.172R,2011MNRAS.415.3693R,2012MNRAS.420..468P,2015A&A...573A..81Y}.
Other non-Band models, such as  synchrotron
model\citep{2011ApJ...741...24B,2014ApJ...784...17B,2014Sci...343...51P},
are also proposed. We do not include the thermal emission in the
analysis, because the thermal emission with a single temperature
is usually found in the careful time-resolved analysis, while our
searching for LAT spectral cutoff needs to be done in the
time-integrated spectrum in order to have enough LAT photons.
Furthermore, we also find that, when a thermal component can be
adequately added in a time-integrated GRB spectrum, the LAT cutoff
energy remains unchanged (although it changes the peak energy of
the Band component). For some non-Band models, such as the
Power-law(PL), smoothly broken power law model(SBPL), Comptonized
model (Comp)\citep{2012ApJS..199...19G,2014ApJS..211...12G}, we
found that they do not improve the fit over the Band model
significantly for the bursts in our sample. For these reasons, and
considering that Band model is  a widely used phenomenological
model, we use the Band model as the primary model in our
analysis}\footnote { There are 3 out of 28 GBM+LAT GRBs(GRB
090531B, 100728A and 110328B) in the first \textsl{Fermi}-LAT GRB
catalog (three years data)(\citet{2013ApJS..209...11A}), which
could be best fitted by a single Comptonized model in the GBM
interval($T_{90}$). But these GRBs would not be detected
significantly in the LAT range during $T_{90}$ with TS smaller
than 9.}. Meanwhile an extra power-law component (sometimes with
high energy cutoff )was also found in some LAT-detected GRBs, such
as GRB~090902B, GRB~090510, and
GRB~090926A~\citep{2009ApJ...706L.138A,2010ApJ...716.1178A,2011ApJ...729..114A}.

Thus, we assume the Band function or the Band plus a power-law
function for the fundamental GRB spectral models. To search for
cutoff-like features at the high-energy end, three spectral models
are considered, i.e.,

(a) Band model, that is
\begin{displaymath}
N_{Band}= B(E) = A\left\{ \begin{array}{ll}
(E/100\,\mathrm{keV})^\alpha e^{(-E(2+\alpha)/E_p)} & \textrm{if $E<E_b$}\\
{[(\alpha - \beta) E_p/(100\,\mathrm{keV}(2+\alpha))]}^{(\alpha-\beta)} \\
e^{\beta-\alpha} (E/100\,\mathrm{keV})^\beta & \textrm{if $E \geq
E_b $}
\end{array} \right.
\end{displaymath}
where $E_b=(\alpha - \beta) E_p/(2+\alpha)$, $\alpha$ is the
photon index at low energy, $\beta$ is the photon index at high
energy and $E_p$ is the peak energy in the $E^2 B(E)$
representation.

(b) BandCut model, the  Band function with an exponential cutoff:
\begin{displaymath}
N_{BandCut}=B(E) e^{-E/E_c},
\end{displaymath}
where $E_c$ is the cutoff energy.

(c) Band+PLcut model, that is the Band function plus a power-law
model with an exponential cutoff
\begin{displaymath}
N_{Band+PLCut}=B(E)+k (E/E_{piv})^\lambda e^{-E/E_c},
\end{displaymath}
where $E_{piv}$ is the pivot energy and $E_c$ is the cutoff
energy.

{To take into account the uncertainties caused by
inter-calibration between the GBM and the LAT, we allow for an
effective area correction during the combined fits. The
calibration constant for the LAT is fixed to one and data from
other detectors are allowed to vary during the
fits~\citep{2013ApJS..209...11A}.} The correction factors
typically have values between 0.9 and 1.1 for the NaI detectors
and between 0.7 and 1.3 for the BGO detectors.

We employ the following three criteria to determine the preferred
spectral model: (1) the goodness of the fitting, which is measured
by the reduced CSTAT (a smaller CSTAT value shows a better fit,
and the difference, $\Delta$CSTAT, is roughly equal to square of
significance of improvement, see~\citet{2011ApJ...729..114A}),
here we claim a significant change with $\Delta$CSTAT larger than
28; (2) the robustness of the model parameters, which is measured
with the errors of the parameters; (3) whether a structure exists
in the residual distribution.

\subsection{Results}
We finally obtain the following results about the joint spectra of \textsl{Fermi} LAT GRBs: 22 GRBs are adequately fitted with the Band function and the rest 6 GRBs show high energy cutoff features: an exponential cutoff from the high energy part of Band component or from extra power law component. The observed spectra with our fitting curve are shown in Figure 1 and the results are reported in Table 2-3.

\subsubsection{Sample fitted by the Band function}
The low energy photon index $\alpha$ of the sub-sample is in the range of $-$1.1 to $-$0.3
with an average value of $-$0.67 and the standard derivation of 0.41,
despite of one GRB with $\alpha$ larger than 0.
The high energy photon index $\beta$ ranges from $-$3.0 to $-$2.0 with an
average value of $-$2.62 and the standard derivation of 0.33. The
peak energy $E_p$ ranges from 70~keV to 900~keV
with an average value of 499~keV and the standard derivation of 387~keV.

{Compared with the recent results of \textsl{Fermi}/GBM catalog:
$E_p=196_{-336}^{+100}$ keV, $\alpha =-1.08_{-0.43}^{+0.44}$ and
$\beta = -2.14_{-0.27}^{+0.37}$ \citep{2014ApJS..211...12G},
\textsl{Fermi}/LAT GRBs (our sample) have  higher peak energy,
harder low energy photon index and softer high energy photon
index} \footnote{ GRB 130427A suffers from pile-up and buffer
saturation effect. We perform spectral analysis with duration
$\sim$138 seconds and find that adding a Power law component can
improve the CSTAT value
significantly~\citep{2014Sci...343...51P,2014Sci...343...42A},
i.e, $\Delta$CSTAT=522. And at its low energy band there maybe
exists a thermal emission
component~\citep{2014Sci...343...51P,2015A&A...573A..81Y}. As
mentioned above, we don't consider it during our fit. For
GRB~140104B, we exclude the events below 50 keV, as its low-energy
part cannot be modeled by any of the aforementioned models.}.

\subsubsection{Sample with high energy spectral cutoff}
Firstly, we reanalyzed the
first \textsl{Fermi}/LAT GRB catalog~\citep{2013ApJS..209...11A}. Only two
GRBs (i.e., GRB 090926A and 100724) are found to show convincing
evidence of  spectral cutoffs or breaks. GRB 090926A and 100724B
can be modelled, respectively, by the Band+PLCut and BandCut
models (see Figure 1).  The results are consistent with the results in
~\citet{2013ApJS..209...11A}.

Among the 28 LAT GRBs detected since August 2011, 6 GRBs are found
to deviate from the Band model with $\Delta$CSTAT$>$28:
GRB~130504C, GRB~130821A, GRB~131231A, GRB~131108A,
GRB~140206B and GRB~141028A. Except for GRB~131108A, BandCut
model is the preferred model with $\Delta$CSTAT large than 28, as
shown in Table 3. Note that, we perform the spectral analysis of
GRB 130821A in the main burst phase, i.e, 3 seconds before and
49 seconds after trigger time~\citep{2013GCN..15261...1J}.
The spectral fits of all the 6 GRBs are
shown in Figure 1.

 We find that the
 Band function cannot fit GRB 131108A's spectrum well, as also noted by ~\citet{2014arXiv1407.0238G}. Following
 our procedures, its time-integrated spectrum can be fitted
 by the Band+PLCut model, with a $\Delta$CSTAT$=$29.5 improvement
 over the Band model. The cutoff energy is found to be at 347.1$\pm$52.8~MeV (see Table 3).
 \citet{2014arXiv1407.0238G} claimed that the
 Band function plus a smoothly broken power function(SBPL) can fit the data well. We test this model and
 find that Band+SBPL has a similar CSTAT value as that of Band+PLCut.
We adopt Band+PLCut as the preferred model since it contains one less parameter than Band+SBPL.

\section{Implications and Discussions}
Including GRB 090926A and GRB 100724B, there are 8 GRBs showing
cutoff-like spectral features in the high-energy emission and the
cutoff energy ranges from $\sim$10 to several hundred MeV.  We note that the cutoffs
obtained for the BandCut model cluster around tens of MeV and no
cutoff above 100 MeV is seen. This may be because the limited
number of photons above 100 MeV does not allow us to distinguish
between the BandCut and the simple Band model statistically if the
break is above 100 MeV. On the other hand, cutoff features can be
discerned more easily if there is an extra hard component (such as
GRB 090926A and GRB 131108A), since the hard power-law component
increases the number of the highest energy photons.

\subsection{Compute the bulk Lorentz factor }
If the spectral break/cutoff $E_c$  is due to $\gamma\gamma$
absorption within the source, one can compute the bulk Lorentz
factor ($ \Gamma $) of the emitting region  by taking
$\tau_{\gamma\gamma}(E_c)=1$. We consider a simple one-zone model
where the photon field in the emitting region is uniform,
isotropic and time independent in the comoving frame {\footnote
{Since  cutoffs  in our sample mostly occur at the high-energy
part of the Band component, it is reasonable to consider the
one-zone model, where high-energy photons come from the same
region as the target photons. For the two-zone model, the
calculation of the Lorentz factor would be
different~\citep{2011ApJ...726L...2Z,2011ApJ...726...89Z}.} }
(see the supporting material for~\citet{2009Sci...323.1688A}).
The target photons that annihilate with photons of
energy $E_c$ should have energy above $E_t =\Gamma^2(m_e
c^2)^2/[E_c(1+z)^2]$, where $z$ is the redshift. These photons
come from the high energy part of the Band function or the extra
power-law component, and their flux can  be, respectively,
parameterized as $f(E)=f(E_0)(E/E_0)^{\beta}$ or
$f(E)=f(E_0)(E/E_0)^{\lambda}$ (in unit $\rm photons/(cm^2keV)$),
where $E_0$ is some reference energy. Considering that photons
with energy $E_c$ collide  with target photons with energy above
$E_t$, we get the $\gamma\gamma$ absorption optical depth in the
comoving frame~\citet{1967PhRv..155.1408G},
\begin{equation}
\tau_{\gamma \gamma }(E'_c)=\int_{-1}^{1} d\mu'
\int_{E'_t}^{E'_{max}}{d E'{n'_{\gamma}(E')} \sigma_{\gamma
\gamma}(E'_c,E',\mu')(1-\mu')W' }
\end{equation}
where $E'_c=(1+z)E_c/\Gamma$ (hereafter the prime represents
quantities in the comoving frame), $\sigma_{\gamma\gamma}$ is the
absorption cross section,  $\mu'=cos\theta'$, $\theta'$ is the
angle between the colliding photon pair,  $W'$ is the shell width,
and $E_{max}$ is the maximum  energy of the target photons. Here
$n'_{\gamma}(E')$ is the number density of target photons in the
comoving frame, which is given by~\citet{2009Sci...323.1688A}
\begin{equation}
n'_{\gamma}(E')=(\frac{d_L}{R})^2 \frac{\Gamma
f(E_0)}{(1+z)^3W'}\left(\frac{E'}{E'_0}\right)^{\beta},
\end{equation}
where $R$ is the radiation radius, and $d_L$ is the luminosity
distance.  Introducing a dimensionless function $F(\beta)$,
\citet{2009Sci...323.1688A}  obtain a simplified expression of
$\tau_{\gamma\gamma}$,
\begin{equation}
\tau_{\gamma \gamma }(E'_c)=\sigma_T (\frac{d_L}{R})^2
\frac{\Gamma E'_0 f(E_0)} {(1+z)^3} (\frac{E'_c E'_0}{m_e^2
c^4})^{-\beta-1}F(\beta),
\end{equation}
where $F(\beta)\approx 0.597(-\beta)^{-2.30}$ for $-2.90 \leq
\beta \leq -1.0 $(Abdo et al. 2009). The relation $R\simeq
\Gamma^2 c \delta t/(1+z)$ is valid for the internal shock model,
where $\delta t$ is the variability time. Setting $\tau_{\gamma
\gamma } (E_{c})= 1$, the Lorentz factor $\Gamma $ is given by
\begin{equation}
\Gamma =[\sigma_T (\frac{d_L}{c \delta t})^2 E_0 f(E_0) F(\beta)
(1+z)^{-2(\beta +1)} (\frac{E_c E_0}{m_e^2
c^4})^{-\beta-1}]^{1/(2(1-\beta ))}.
\end{equation}

One should note that Eq.(3)  is obtained when the upper limit
 of the second integral in Eq.(1), $E_{max}$,  is taken to be
$\infty$, which is valid only when the energy of target photons
that annihilate with $E_c$ is well below the cutoff energy, i.e.,
\begin{equation}
E_{c}\gg \Gamma^2 m_e^2 c^4/[E_c(1+z)^2]
\end{equation}
~\citep{2010ApJ...709..525L,2011ApJ...726...89Z}. This condition
is usually satisfied when the cutoff energy $E_c$ is larger than a
few hundreds MeV. However, for bursts with lower $E_c$, such as
some bursts in our sample, this condition is not satisfied
anymore. For these low $E_c$ bursts, the energy of target photons
should be comparable to $E_c$ (i.e., $E_{c}\ga \Gamma^2 m_e^2
c^4/[E_c(1+z)^2]$), then $\Gamma $ is estimated to
be~\citep{2010ApJ...709..525L}
\begin{equation}
\Gamma \approx \frac{E_{c}}{m_e c^2}(1+z).
\end{equation}

Using the above method, we can now calculate $\Gamma $ for each
burst with  spectral cutoffs. For GRB090926A,
since the time-resolved spectra of the maximum spikes show
spectral cutoffs,  we use the cutoff energy for the time-resolved
spectrum to calculate $\Gamma$. For 4 GRBs that do not have
redshift measurements, we assume redshifts of $z=1$ for them. The
variability time $\delta t =0.1 {\rm s}$  is adopted for
GRB131108A, based on the 2 ms resolution lightcurve of the bright
NaI detector. For GRB090926A, $\delta t =0.15 {\rm s}$  is
adopted according to ~\citep{2011ApJ...729..114A}. We first use Eq. (4) to
calculate $\Gamma$, and then check whether Eq.(5) is satisfied. We
find that, only two bursts (GRB131108A and GRB090926A),  which
have relatively larger $E_c$, satisfy this condition.  The other
seven bursts all have $E_c\la 100 {\rm MeV}$ and Eq.5 is not
satisfied for them, so their Lorentz factor $\Gamma $  are
calculated with Eq.(6). We note that this estimate of $\Gamma$
suffers from less assumptions, as they are independent of the
internal shock model assumption and the estimate of the
variability timescale. The results of $\Gamma$ are presented in
Table 5. The values of $\Gamma$ in our sample range from 90 to
900, providing direct evidence that GRBs are powered by
ultra-relativistic outflow.

{It has been usually suggested that, even if no spectral cutoff is
measured, the observed highest energy photon  can be used to place
a lower limit on the bulk Lorentz factor, assuming that the
absorption optical depth $\tau_{\gamma\gamma}(E_{\rm max})\la1$
for the maximum energy
photon(~\citep{1991ApJ...373..277K,1993A&AS...97...59F,1995ApJ...453..583W,1997ApJ...491..663B,2012MNRAS.421..525H}).
However, from our sample that have measured cutoffs,  one can see
that the absorption optical depth equals unity for the cutoff
energy (i.e. $\tau_{\gamma\gamma}(E_c)=1$) and   the absorption
optical depth for the maximum energy photon is larger than unity
(i.e. $\tau_{\gamma\gamma}(E_{\rm max})>1$). For this reason, the
usual approach that uses  $\tau_{\gamma\gamma}(E_{\rm max})\la1$
for the highest energy photon to estimate the lower limits on the
bulk Lorentz factors is inaccurate.}

{Another method for estimating the bulk Lorentz factors  is from
the peak in the early optical afterglow light curve, assuming that
this peak is caused by the afterglow onset, at which the jet is
decelerated~\citep{1999ApJ...520..641S,2010ApJ...725.2209L,2011ApJ...738..138R,2014ApJ...782....5H}.
Although most of the LAT bursts do not have early optical
afterglow data, such estimate may be possible in some cases.  The
Lorentz factors can also be determined from the thermal component
in the prompt emission, assuming it comes from the fireball
photosphere~\citep{2007ApJ...664L...1P}. But this method depends
on the unknown composition of GRBs outflow and the efficiency of
dissipation mechanism responsible for the non-thermal
component~\citep{2014ApJ...795..155P,2014arXiv1409.3584G}.}

\subsection{Correlations in $\Gamma-L_{\gamma,iso}$ and $\Gamma-E_{\gamma,iso}$}
There have been suggestions that the Lorentz factors correlate
with other quantities of the GRB jets, such as the isotropic
gamma-ray luminosity or
energy~\citep{2010ApJ...725.2209L,2012ApJ...751...49L,2012MNRAS.420..483G}.
The Lorentz factors in all the references are determined from the
afterglow onset time,  at which the jet is decelerated, so the
values depends on the details of the dynamics and the
circumburst environment.  The Lorentz factors determined through
the absorption cutoff in high-energy photons is more
straightforward and reliable. We test the relation
$\Gamma-L_{\gamma,iso}$ and $\Gamma-E_{\gamma,iso}$ using our
sample, where $L_{\gamma,iso}$ is the averaged, isotropic
gamma-ray luminosity in 10-1000~keV and $E_{\gamma,iso}$ is the
isotropic gamma-ray energy in 10-1000~keV. The results are shown
in Fig.2. We find the relation
\begin{equation}
\Gamma = 10^{1.65\pm0.20} L_{\gamma,iso,51}^{0.52\pm0.13},
\end{equation}
with a Pearson correlation coefficient of $r=0.844$ and null
hypothesis probability of 0.008, which indicates a tight positive
correlation. Removing the 4 GRBs that do not have redshift
measurements, we  examine whether the correlation remains.
Although the sample gets smaller, we find that a correlation
between $\Gamma$ and $L_{\gamma,iso}$ is still keep. Similarly, for the
8 GRBs, we find that
\begin{equation}
\Gamma = 10^{1.01\pm0.56} E_{\gamma,iso,52}^{0.88\pm0.35},
\end{equation}
with a Pearson correlation coefficient of $r=0.707$ and null
hypothesis probability of 0.050.  Although our results generally
agree with earlier suggestions that more powerful GRBs move
faster, the correlation slopes are  different. We note that the
number of GRBs in our sample is  limited and the correlation
remains to be tested with a large sample in future.

GRB jets are accelerated at the early stage while the  internal
energy of the fireball is gradually converted to the kinetic
energy. After the acceleration, the jet is expected to have the
Lorentz factor equal to the the initial dimensionless entropy
$\eta = L_0/( \dot{M} c^2)$, where  $L_0$ and $\dot{M}$ are
respectively the total  energy and mass outflow rates. Considering
the relation in Eq.7,  the mass outflow rates should follow that
$\dot{M}\propto {L_{\gamma,iso}^{0.48\pm0.13}}$ (assuming that
$L_{\gamma,iso}\propto L_0$). This put useful constraint on any
central engine models for GRBs.

\subsection{Summary}
We perform a complete analysis of the LAT-detected GRBs since
August 2011, i.e. the bursts that are not included in the first
\textsl{Fermi}/LAT GRB catalog  (Ackermann  et al. 2013). Our aim
is to search for cutoff-like spectral feature in  the high-energy
gamma-ray emission, as has been seen in GRB 090926A. We find 6
GRBs showing such spectral features, with the cutoff energies
ranging from $\sim10$ to $\sim500$ MeV.  Assuming a simple
one-zone model for the MeV-GeV emission, we compute the bulk
Lorentz factors of the emitting region of these bursts. {Motivated
by earlier suggestion that the Lorentz factors may correlate with
other burst quantities, such as the isotropic gamma-ray luminosity
or energy
\citep{2010ApJ...725.2209L,2012ApJ...751...49L,2012MNRAS.420..483G},
we test these relations  with our sample.} It is found that the
Lorentz factors are well correlated with the isotropic gamma-ray
luminosity of the bursts, suggesting that more powerful GRB
outflow move faster.

\acknowledgments \acknowledgments We thank the referee for the
constructive report, and thank Zhuo Li, Xue-Feng Wu, and Hoi-Fung Yu for useful
discussions. This work has made use of data and software provided
by the Fermi Science Support Center.
 This work is supported by the 973 program
under grant 2014CB845800, the NSFC under grants 11273016 and
11033002, and the Excellent Youth Foundation of Jiangsu Province
(BK2012011). PHT is supported by the One Hundred Talents Program
of the Sun Yat-Sen University.

{}

\clearpage

\begin{deluxetable}{lccccccccr}
\tablenum{1} \tablewidth{0pt} \topmargin 0.0cm \evensidemargin =
0mm \oddsidemargin = 0mm \tabletypesize{\scriptsize}
\tablecaption{Properties of \textsl{Fermi}/LAT GRBs from Aug 2011 to Oct
2014} \tablehead{ \colhead{GRB name}&
\colhead{Class\tablenotemark{a}} &
\colhead{T$_{90}$\tablenotemark{b}} &
\colhead{T$_{90}^S$\tablenotemark{b}} &
\colhead{$\theta$\tablenotemark{c}}&
\colhead{RA   \tablenotemark{d}} &
\colhead{Decl.\tablenotemark{d}} &
\colhead{LLE\&GBM\tablenotemark{e}} &
\colhead{Ref\tablenotemark{f}}\\
&  & (s) & (s) &(deg)&(deg) & (deg) & &  }
\startdata
120107A &   L   &   23.04   &   0.064   &   56  &   246.4   &   -69.93  &   b0,n6,n7    &   1   \\
120316A &   L   &   26.624  &   1.536   &   9   &   57.97   &   -56.46  &   LLE,b0,n0,n1    &   2   \\
120709A &   L   &   27.328  &   -0.128  &   22  &   318.41  &   -50.03  &   LLE,b1,n6,n7,n9 &   3   \\
120830A &   S   &   1.28    &   -0.384  &   38  &   88.42   &   28.81   &   b0,n0,n1,n3 &   4   \\
130327B &   L   &   31.233  &   2.048   &   47  &   218.09  &   -69.51  &   b0,n0,n1    &   5   \\
130427A &   L   &   138.242 &   4.096   &   48  &   173.15  &   27.71   &   LLE,b1,n9,n10   &   6   \\
130502B &   L   &   24.32   &   7.168   &   47  &   66.65   &   71.08   &   b1,n6,n7    &   7   \\
130504C &   L   &   73.217  &   8.704   &   47  &   91.72   &   3.85    &   LLE,b0,n2,n9    &   8   \\
130518A &   L   &   48.577  &   9.92    &   43  &   355.81  &   47.64   &   LLE,b1,n3,n7    &   9   \\
130821A &   L   &   87.041  &   3.584   &   37  &   314.1   &   -12 &   LLE,b1,n6,n9    &   10  \\
130828A &   L   &   136.45  &   13.312  &   40  &   259.83  &   28  &   b0,n0,n3    &   11  \\
131014A &   L   &   3.2 &   0.96    &   71.9    &   100.5   &   -19.1   &   LLE,b1,n9,na,nb &   12  \\
131018B &   L   &   39.936  &   -1.024  &   12  &   304.41  &   23.11   &   b1,n6,n7    &   13  \\
131029A &   L   &   104.449 &   1.024   &   6   &   200.79  &   48.3    &   b0,n3,n5    &   14  \\
131108A &   L   &   18.496  &   0.448   &   27  &   156.47  &   9.9 &   LLE,b1,n3,n6    &   15  \\
131209A &   L   &   13.568  &   2.816   &   20  &   136.5   &   -33.2   &   b1,n6,n7    &   16  \\
131231A &   L   &   31.232  &   13.312  &   40  &   10.59   &   -1.85   &   LLE,b0,n0,n3    &   17  \\
140102A &   L   &   3.648   &   0.448   &   47  &   211.88  &   1.36    &   LLE,b1,n6,n7,n9,nb  &   18  \\
140104B &   L   &   188.417 &   9.216   &   25  &   218.81  &   -8.9    &   b1,n6,n7    &   19  \\
140110A &   L   &   9.472   &   -0.256  &   30  &   28.9    &   -36.26  &   LLE,b1,n6,n7,n9 &   20  \\
140206B &   L   &   116.738 &   8.256   &   45  &   315.26  &   -8.51   &   LLE,b0,n0,n1,n3 &   21  \\
140323A &   L   &   111.426 &   5.056   &   31  &   356.46  &   -79.87  &   b0,n0,n1    &   22  \\
140402A &   S   &   0.32    &   -0.128  &   13  &   207.47  &   5.87    &   b0,n1,n3    &   23  \\
140523A &   L   &   19.2    &   0.576   &   60  &   133.3   &   24.95   &   b0,n3,n4    &   24  \\
140619B &   S   &   2.816   &   -0.256  &   32  &   132.68  &   -9.66   &   LLE,b1,n6,n9    &   25  \\
140723A &   L   &   56.32   &   0   &   55  &   210.63  &   -3.73   &   b1,n9,na    &   26  \\
140729A &   L   &   55.553  &   0.512   &   26.2    &   193.95  &   15.35   &   LLE,b1,n6,n8,n9 &   27  \\
141028A &   L   &   31.489  &   6.656   &   25  &   322.7   &   -0.28   &   LLE,b1,n6,n7,n9 &   28  \\
090926A &   L   &   13.76   &   2.176   &   48.1    &   353.4 & -66.32   &   LLE,b1,n3,n6,n7 &   29  \\
100724B &   L   &   114.69  &   8.192   &   48.9    &   119.89  &   76.55   &   LLE,b0,n0,n1    &   29
\enddata
\tablenotetext{a}{{\rm L} means long burst and S means
short burst.} \tablenotetext{b}{GBM T90 duration and the start
time from GBM trigger time cited from GBM catalog, i.e,
~\citet{2012ApJS..199...18P,2012ApJS..199...19G,2014ApJS..211...12G,2014ApJS..211...13V}.}
\tablenotetext{c}{Off-axis angle at the trigger time derived from
reference in the ninth collum.} \tablenotetext{d}{{\rm LAT} position
from reference in the ninth collum.} \tablenotetext{e}{{\rm LLE}
represents the publicly LAT Low-Energy data; others are the GBM
 detectors we used for spectral analysis.}
\tablenotetext{f}{\textbf{Reference:}{ 1:
Zheng, W. \& Akerlof, C. (2012); McBreen, S. (2012); 2: Vianello,
G. et al. (2012); 3: Kocevski, D. et al. (2012); Guiriec, S. et
al. (2012); 4: Vianello, G. et al. (2012); Tierney, D. (2012); 5:
Ohno, M. et al. (2013); Chaplin, V.\& Fitzpatrick, G. (2013); 6:
Zhu, S. et al. (2013); von Kienlin, A. (2013); 7: Kocevski, D. et
al. (2012); von Kienlin, A. \& Younes, G. (2013); 8: Kocevski, D.
et al. (2013); Burgess, J. M. et al. (2013); 9: Omodei, N. \&
McEnery, J. (2013); Xiong, S. (2013); 10: Kocevski, D. et al.
(2013); Jenke, P. (2013b); 11: Vianello, G. \& Sonbas, E. (2013);
Collazzi, A. C. (2013); 12: Desiante, R. et al. (2013);
Fitzpatrick, G. \& Xiong, S.(2013); 13: Vianello, G. et al.
(2013); Zhang, B.-B. (2013); 14: Racusin, J. L. et al. (2013); von
Kienlin, A.\& Jenke, P. (2013); 15: Racusin, J. L. et al. (2013);
Younes, G. (2013); 16: Vianello, G.\& Omodei, N. (2013); von
Kienlin, A.\& Meegan, C. (2013); 17: Sonbas, E. et al. (2013);
Jenke, P.\& Xiong, S. (2014); 18: Sonbas, E. et al. (2014); Zhang,
B.-B.\& Bhat, N. (2014); 19: Vianello, G. et al. (2014); Xiong, S.
(2014); 20: Bissaldi, E. et al. (2014); von Kienlin, A.\&
Connaughton, V. (2014); 21: Bissaldi, E. et al. (2014); von
Kienlin, A. (2014); 22: Vianello, G. et al. (2014); Yu, H.-F.\&
von Kienlin, A. (2014); 23: Bissaldi, E. et al. (2014): Jenke, P.
A.\& Yu, H.-F. (2014); 24: Vianello, G. et al. (2014); von
Kienlin, A.\& Connaughton, V. (2014); 25: Kocevski, D. et al.
(2014); Connaughton, V. et al. (2014); 26: Bissaldi, E. et al.
(2014); Burns, E. (2014); 27: Arimoto, M.\& Bissaldi, E. ((2014);
Stanbro, M. (2014); 28: Bissaldi, E.:et al. (2014);Roberts, O. J.
(2014). 29: Ackermann et al. (2013)}}
\end{deluxetable}

\begin{deluxetable}{lcrrcr}
\tablenum{2} \tablewidth{0pt} \topmargin 0.0cm \evensidemargin =
0mm \oddsidemargin = 0mm \tabletypesize{\scriptsize}
\tablecaption{Joint spectral fits of the sample modelled
by the Band function} \tablehead{
\colhead{GRB name} & \colhead{model} & \colhead{$\alpha$ } &
\colhead{$\beta$ } & \colhead{$E_p$(keV) } & \colhead{CSTAT/DOF  }
} \startdata
120107A &   Band    &   -1.19   $\pm$   0.08    &   -2.39   $\pm$   0.11    &   275.2   $\pm$   59.7    &   396/381 \\
120316A &   Band    &   -0.74   $\pm$   0.03    &   -2.71   $\pm$   0.11    &   421.4   $\pm$   14.4    &   17216/383   \\
120709A &   Band    &   -1.06   $\pm$   0.04    &   -2.59   $\pm$   0.07    &   423.1   $\pm$   39.2    &   715/510 \\
120830A &   Band    &   -0.13   $\pm$   0.11    &   -2.63   $\pm$   0.11    &   887.7   $\pm$   103.0   &   576.9/505   \\
130327B &   Band    &   -0.64   $\pm$   0.02    &   -2.74   $\pm$   0.09    &   327 $\pm$   8.2 &   607.7/383   \\
130427A &   Band+PL &   -0.87   $\pm$   0.01    &   -2.83   $\pm$   0.01    &   900.1   $\pm$   7.0 &   1146/370    \\
130502B &   Band    &   -0.51   $\pm$   0.01    &   -2.61   $\pm$   0.03    &   280.6   $\pm$   3.6 &   594/370 \\
130518A &   Band    &   -0.89   $\pm$   0.01    &   -2.72   $\pm$   0.04    &   400 $\pm$   10.3    &   676.8/385   \\
130828A &   Band    &   -1.12   $\pm$   0.11    &   -2.45   $\pm$   0.06    &   243.5   $\pm$   21.7    &   549/326 \\
131014A &   Band    &   -0.21   $\pm$   0.01    &   -2.62   $\pm$   0.02    &   308.5   $\pm$   2.7 &   990/487 \\
131018B &   Band    &   -0.20   $\pm$   0.41    &   -3.77   $\pm$   2.61    &   77.7    $\pm$   10.8    &   608.2/381   \\
131029A &   Band    &   -0.98   $\pm$   0.05    &   -2.32   $\pm$   0.05    &   230.2   $\pm$   20.6    &   575/381 \\
131209A &   Band    &   -0.34   $\pm$   0.05    &   -2.97   $\pm$   0.36    &   281.3   $\pm$   13.9    &   401/381 \\
140102A &   Band    &   -0.75   $\pm$   0.02    &   -2.58   $\pm$   0.04    &   182.1   $\pm$   4.3 &   808/632 \\
140104B\tablenotemark{a}    &   Band    &   -0.68   $\pm$   0.16    &   -3.00(fixed)    &   218.8   $\pm$   14.7    &   840.9/324    \\
140110A &   Band    &   -0.72   $\pm$   0.06    &   -2.53   $\pm$   0.07    &   1431    $\pm$   266 &   900/511 \\
140323A &   Band    &   -0.99   $\pm$   0.03    &   -2.41   $\pm$   0.06    &   143.2   $\pm$   6.8 &   3185/383    \\
140402A &   Band    &   0.49    $\pm$   0.62    &   -2.28   $\pm$   0.1 &   715.2   $\pm$   202 &   397/382 \\
140523A &   Band    &   -0.94   $\pm$   0.01    &   -2.62   $\pm$   0.06    &   243.2   $\pm$   5.8 &   549/380 \\
140619B &   Band    &   -0.28   $\pm$   0.32    &   -2.14   $\pm$   0.05    &   680.6   $\pm$   215 &   419/397 \\
140723A &   Band    &   -1.14   $\pm$   0.05    &   -2.34   $\pm$   0.07    &   1383    $\pm$   460 &   822.1/380   \\
140729A &   Band    &   -0.86   $\pm$   0.06    &   -2.74   $\pm$   0.11    &   929.4   $\pm$   155 &   1813/504
\enddata
\tablenotetext{a}{In this fit, the
energy range of NaI starts from >$\sim$50 keV.}
\end{deluxetable}

\begin{deluxetable}{llrrcccrcr}
\tablenum{3} \tablewidth{0pt} \topmargin 0.0cm \evensidemargin =
0mm \oddsidemargin = 0mm \tabletypesize{\scriptsize}
\tablecaption{Joint spectral fits for the sample with
high energy cutoffs} \tablehead{ \colhead{GRB name} &
\colhead{model} & \colhead{$\alpha$ } & \colhead{$\beta$ } &
\colhead{$E_p$(keV) } & \colhead{$\lambda$ } & \colhead{$E_c$(MeV)
} & \colhead{CSTAT/DOF } & \colhead{$\Delta$CSTAT  } } \startdata
090926A &   Band+PLCut  &   -0.50   $\pm$   0.03    &   -2.54   $\pm$   0.03    &   269.8   $\pm$   3.7 &   -1.78   $\pm$   0.02     &   550.2  $\pm$   91.5    &   983.1/486   &   135.8   \\
... &   Band    &   -0.71   $\pm$   0.01    &   -2.31   $\pm$   0.01    &   285.3   $\pm$   3.2 &   --          &   --          &    1118.9/489 &   --  \\
100724B &   BandCut &   -0.71   $\pm$   0.01    &   -2.08   $\pm$   0.01    &   354.5   $\pm$   1.5 &   --          &   42.4     $\pm$   4.0    &   1202.3/389  &   342.8   \\
... &   Band    &   -0.77   $\pm$   0.01    &   -2.43   $\pm$   0.01    &   417.9   $\pm$   6.6 &   --          &   --          &    1545.1/390 &   --  \\
130504C &   BandCut &   -1.21   $\pm$   0.01    &   -2.03   $\pm$   0.01    &   619.6   $\pm$   7.8 &   --          &   22.2     $\pm$   6.3    &   740.3/389   &   70.2    \\
... &   Band    &   -1.23   $\pm$   0.01    &   -2.66   $\pm$   0.03    &   722.1   $\pm$   30.7    &   --          &   --           &   810.5/390  &   --  \\
130821A &   BandCut &   -1.04   $\pm$   0.01    &   -2.12   $\pm$   0.02    &   297.6   $\pm$   2.9 &   --          &   13.3     $\pm$   7.3    &   793.7/388   &   40.7    \\
... &   Band    &   -1.08   $\pm$   0.01    &   -2.78   $\pm$   0.05    &   341.9   $\pm$   10.8    &   --          &   --           &   834.4/389  &   --  \\
131108A &   Band+PLCut  &   -0.69   $\pm$   0.09    &   -2.59   $\pm$   0.16    &   291.5   $\pm$   15.8    &   -1.69   $\pm$    0.04   &    347.1  $\pm$   52.8    &   411.4/385   &   29.5    \\
... &   Band    &   -0.88   $\pm$   0.03    &   -2.16   $\pm$   0.01    &   308.5   $\pm$   14.6    &   --          &   --           &   440.9/388  &   --  \\
131231A &   BandCut &   -1.21   $\pm$   0.01    &   -2.43   $\pm$   0.01    &   205.9   $\pm$   0.8 &   --          &   61.6     $\pm$   22.5   &   1175.8/380  &   31  \\
... &   Band    &   -1.22   $\pm$   0.01    &   -2.62   $\pm$   0.02    &   214.3   $\pm$   3.1 &   --          &   --          &    1206.8/381 &   --  \\
140206B &   BandCut &   -1.14   $\pm$   0.01    &   -2.03   $\pm$   0.01    &   241.2   $\pm$   1.9 &   --          &   50.1     $\pm$   6.8    &   2392.0/511  &   108.4   \\
... &   Band    &   -1.17   $\pm$   0.01    &   -2.34   $\pm$   0.01    &   276.6   $\pm$   7.4 &   --          &   --          &    2500.4/512 &   --  \\
141028A &   BandCut &   -0.83   $\pm$   0.01    &   -2.05   $\pm$   0.01    &   288.6   $\pm$   2.9 &   --          &   53.2     $\pm$   7.3    &   1101.3/510  &   92.8    \\
... &   Band    &   -0.86   $\pm$   0.02    &   -2.37   $\pm$   0.02    &   316.3   $\pm$   11.7    &   --          &   --           &   1194.1/511 &   --\\
\hline
090926A-a   &   Band+PLCut  &   -0.91   $\pm$   0.09    &   -2.66   $\pm$   0.37    &   226.2   $\pm$   9.2 &   -1.69   $\pm$    0.03   &   350.7   $\pm$   41.3    &   492.4/479   &   68.1    \\
... &   Band    &   -1.01   $\pm$   0.03    &   -2.12   $\pm$   0.01    &   231.4   $\pm$   10.1    &   --  &   --  &    560.5/482   &  --
\enddata
 \tablecomments{The top panel are the results for the  time-integrated
 spectra. The bottom panel are the results for the  time-resolved spectra of
 GRB 090926A in the interval of [9.79s,10.50s].}
\end{deluxetable}

\begin{deluxetable}{lcccrrrc}
\tablenum{4} \tablewidth{0pt} \topmargin 0.0cm \evensidemargin =
0mm \oddsidemargin = 0mm \tabletypesize{\scriptsize}
\tablecaption{Burst parameters and the derived Lorentz factor of
GRBs}
\tablehead{
\colhead{GRB name} &
\colhead{$E_c$(~MeV)} &
\colhead{$z$\tablenotemark{a}} &
\colhead{$L_{\gamma,iso}$ \tablenotemark{b}} &
\colhead{$E_{\gamma,iso}$\tablenotemark{c} } &
\colhead{$\Gamma$ \tablenotemark{d}} &
\colhead{$z$ Ref \tablenotemark{e} }
} \startdata
100724B &       42.4    $\pm$   4   &   --  &   10.1    $\pm$   0.1 &   58.2    $\pm$   0.6 &   165.9   $\pm$   15.6    &   --   \\
130504C &       22.2    $\pm$   6.3 &   --  &   8.9 $\pm$   0.1 &   32.7    $\pm$   0.3 &   86.7    $\pm$   24.5    &   --  \\
130821A &       13.3    $\pm$   7.3 &   --  &   7.1 $\pm$   0.1 &   18.4    $\pm$   0.3 &   52.1    $\pm$   28.6    &   --  \\
131231A &       61.6    $\pm$   22.5    &   0.642   &   8.7 $\pm$   0.1 &   16.5    $\pm$   0.1 &   197.9   $\pm$   72.3    &    (1)    \\
140206B &       50.1    $\pm$   6.8 &   --  &   5.2 $\pm$   0.1 &   30.2    $\pm$   0.4 &   196.1   $\pm$   26.6    &   --  \\
141028A &       53.2    $\pm$   7.3 &   2.332   &   57.7    $\pm$   1.3 &   54.5    $\pm$   1.2 &   346.9   $\pm$   47.6    &    (2)    \\
131108A &       347.1   $\pm$   52.8    &   2.4 &   90.7    $\pm$   1.2 &   49.4    $\pm$   0.7 &   734.4   $\pm$   111.7   &    (3)    \\
090926A &       350.7   $\pm$   41.3    &   2.1 &   365.1   $\pm$   13.7    &   215.1   $\pm$   8.1 &   748.3   $\pm$   88.1    &    (4)
\enddata
\tablenotetext{a}{Redshifts of GRBs. Bursts
with no redshift measurements are assumed to have $z$ = 1.}
\tablenotetext{b}{Isotropic gamma-ray luminosity in 10-1000~keV
obtained from the best fit of each GRB in unit of $10^{50}$ erg s$^{-1}$. }
\tablenotetext{c}{Isotropic gamma-ray energy in 10-1000~keV
obtained from the best fit of each GRB in unit of $10^{52}$ erg.}
\tablenotetext{d}{The bulk Lorentz factor. }
\tablenotetext{e}{(1) Xu, D. et al. (2014b), Cucchiara, A. (2014a);
(2) de Ugarte Postigo, A. et al. (2013), Xu, D. et al. (2013);
(3) Xu, D. et al. (2014a); (4) Ackermann et al.(2011).}
\end{deluxetable}

        \begin{figure}
        \figurenum{1}
    \epsscale{.4}
    \plotone{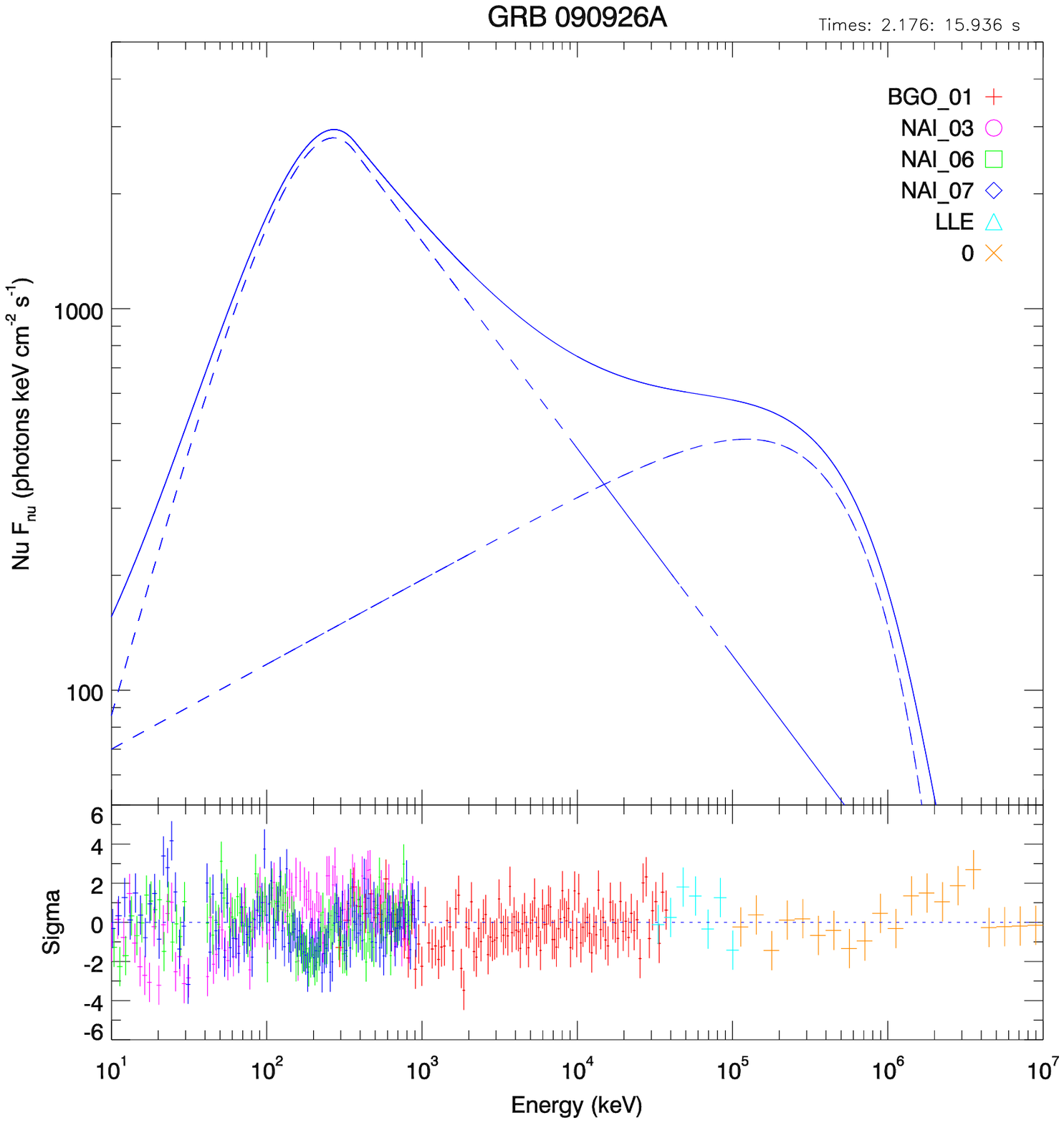}
    \plotone{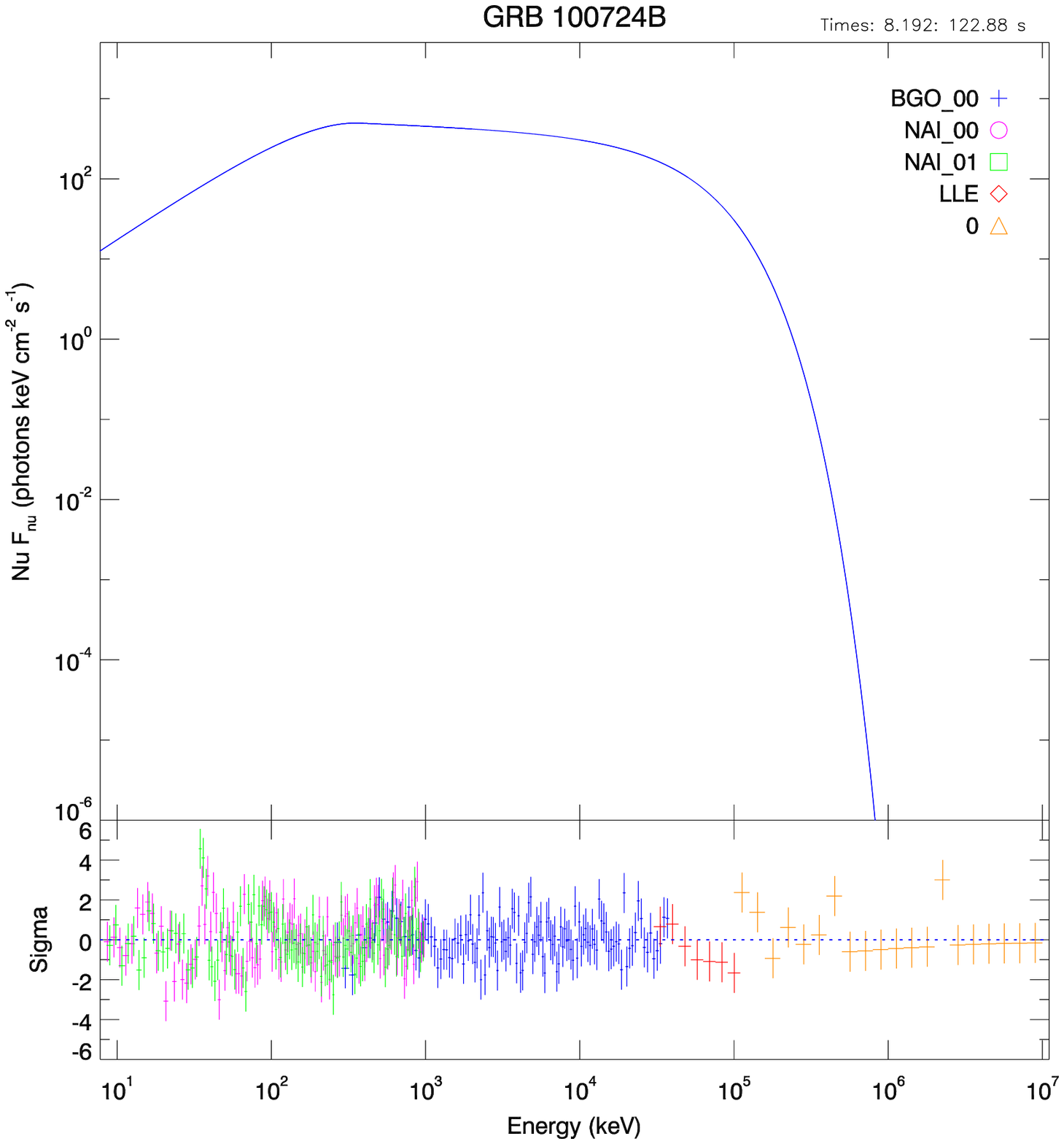}
    \plotone{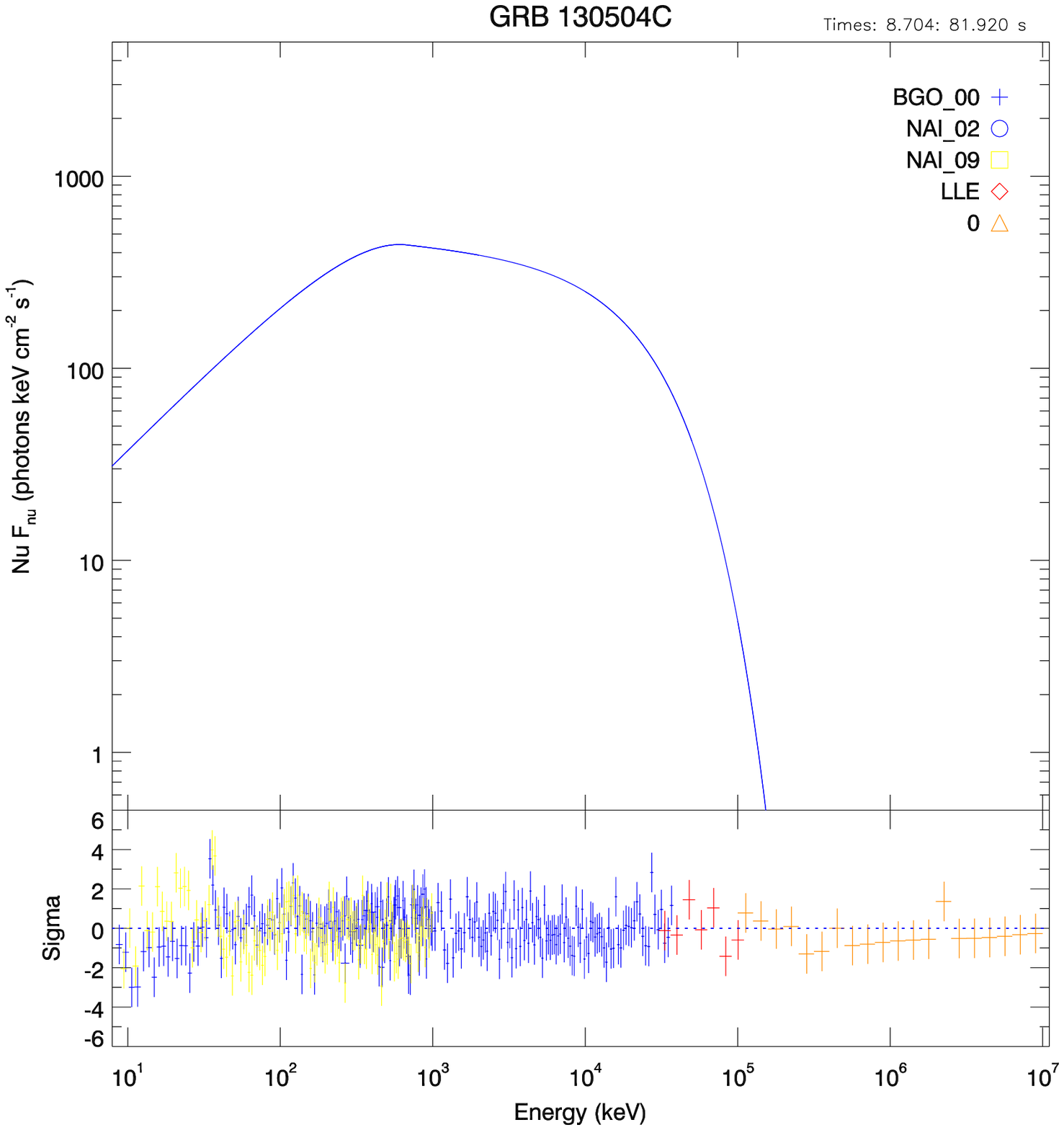}
    \plotone{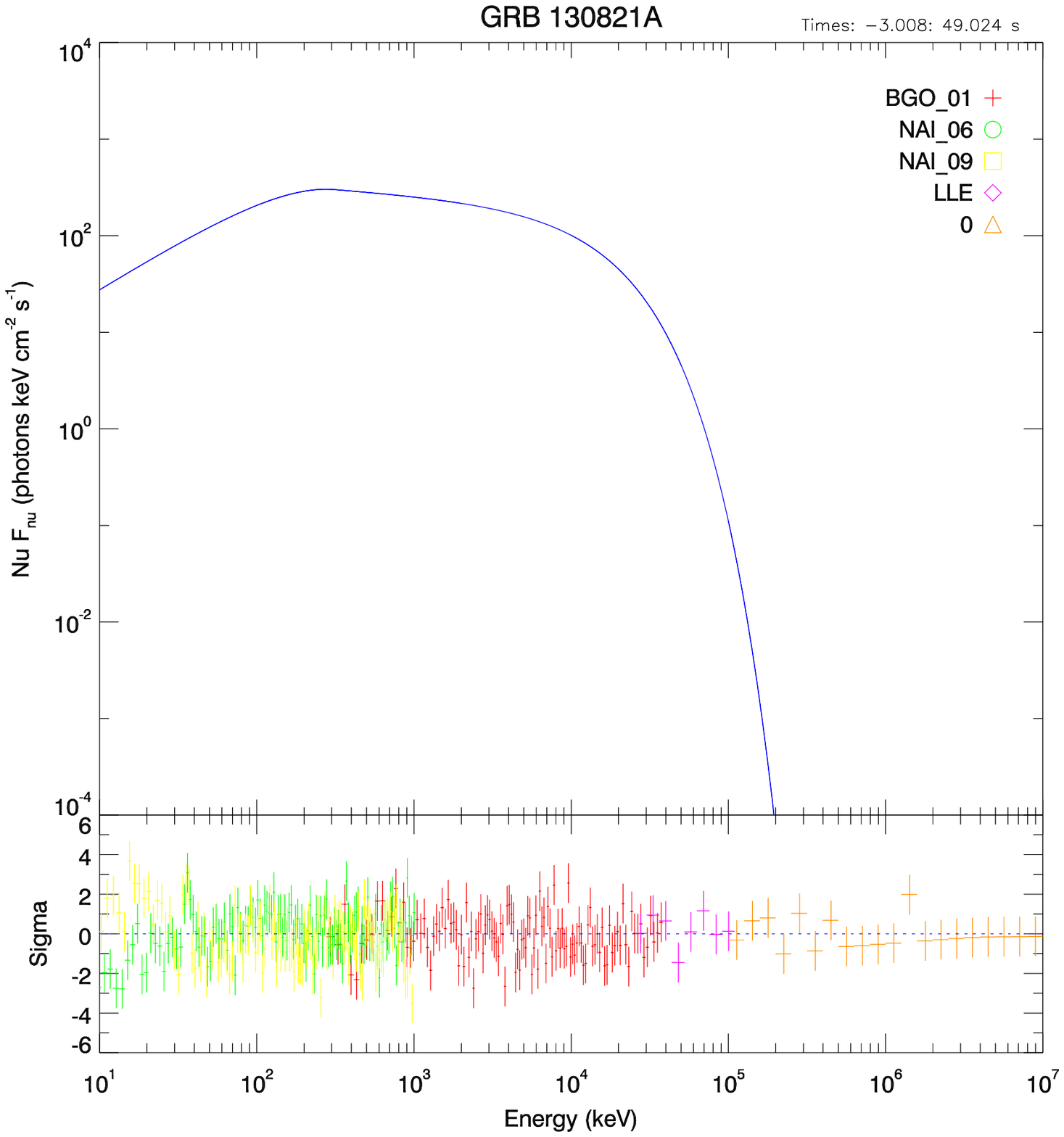}
    \plotone{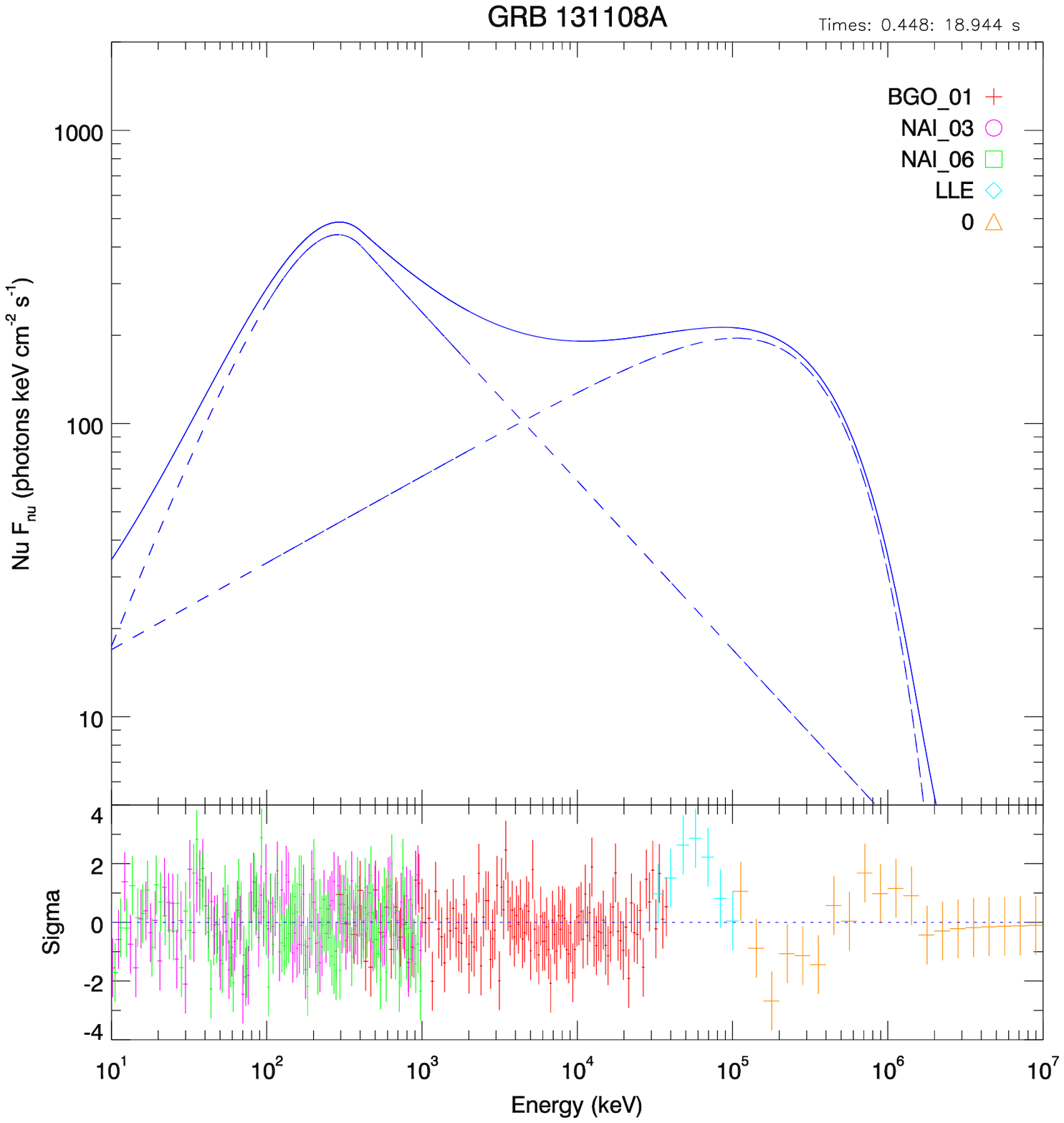}
    \plotone{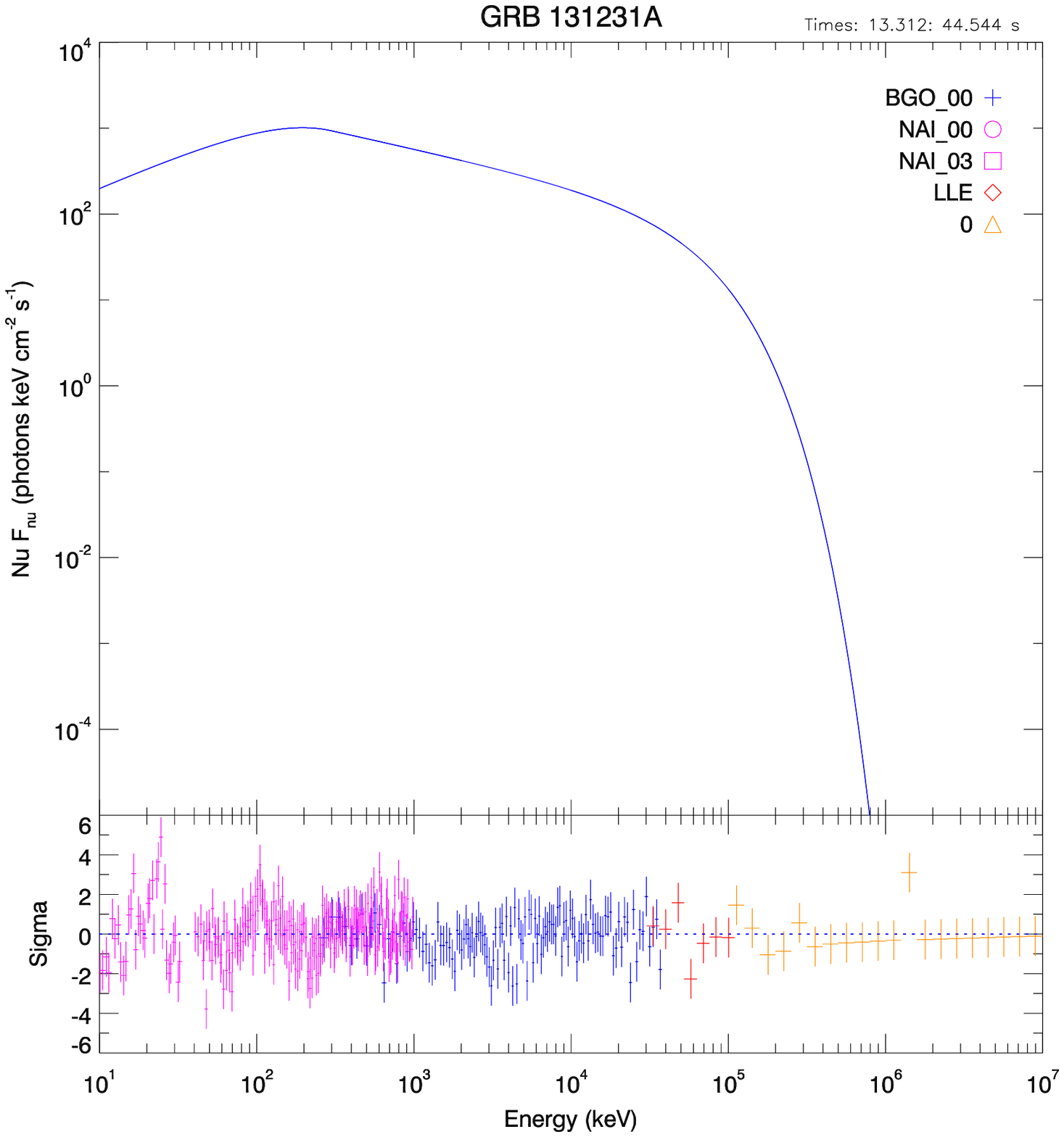}
   \caption{Spectral fits and residuals of the time integrated emission and the best-fit model of 8 GRBs showing high energy cutoffs. The top panels show $\nu F_\nu$ spectra and the bottom panels show the residuals of the fit. The "0" represents the LAT data.}
   \label{gbmlatseds}
        \end{figure}

            \begin{figure}
            \figurenum{1}
    \epsscale{.4}
    \plotone{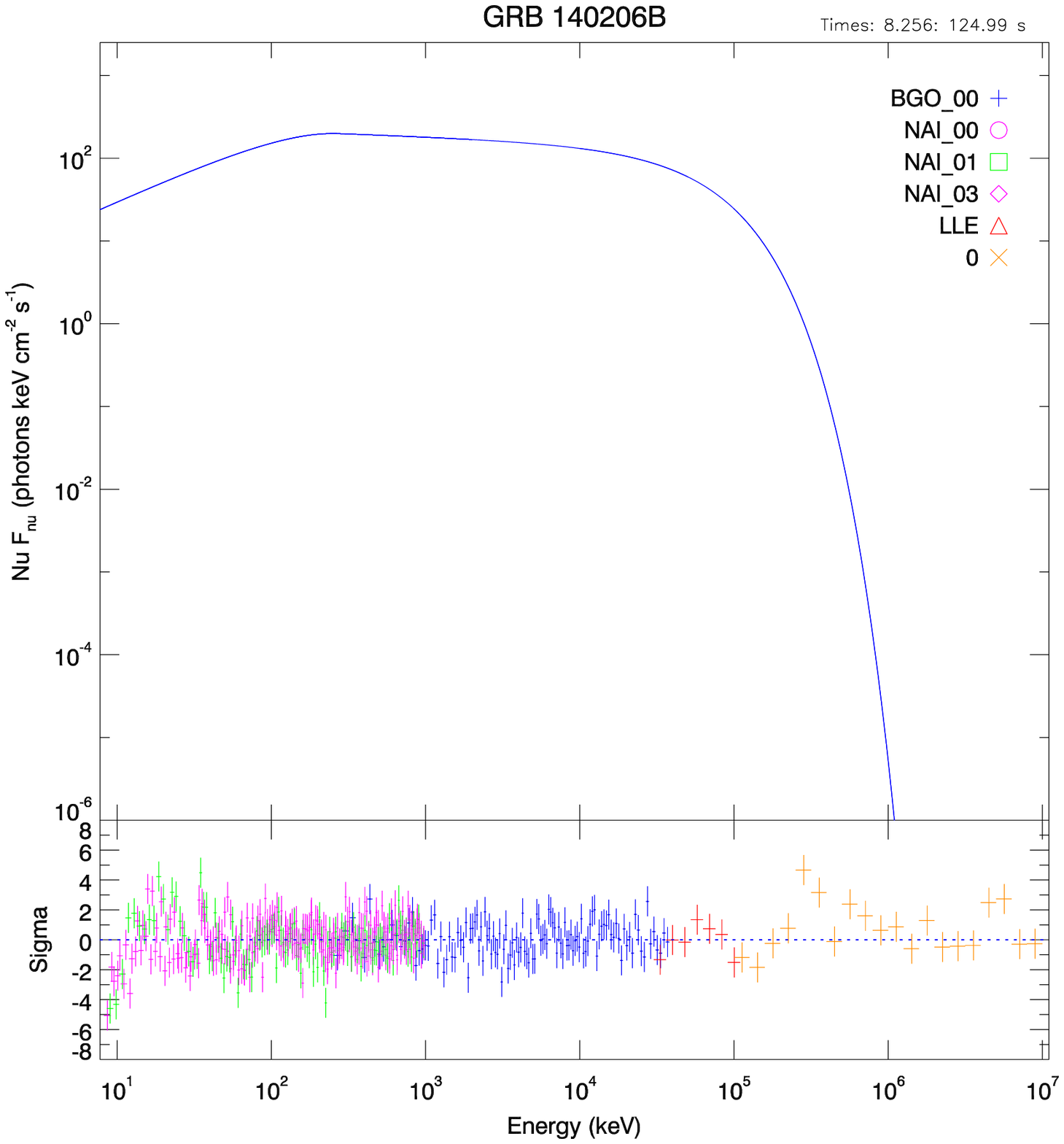}
    \plotone{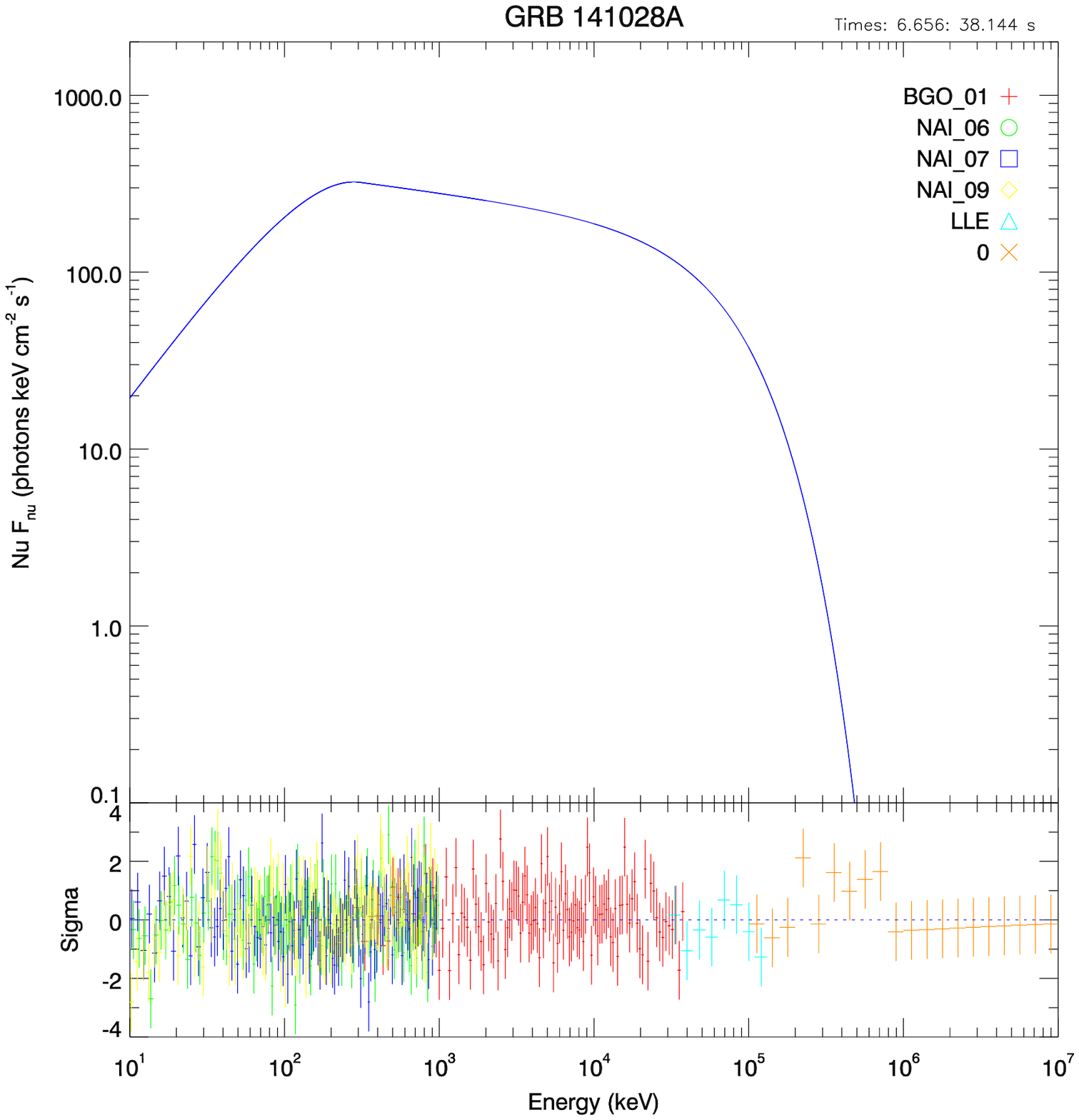}
   \caption{--continued}
   \label{gbmlatseds}
        \end{figure}

 \begin{figure}
  \figurenum{2}
    \epsscale{.5}
       \plotone{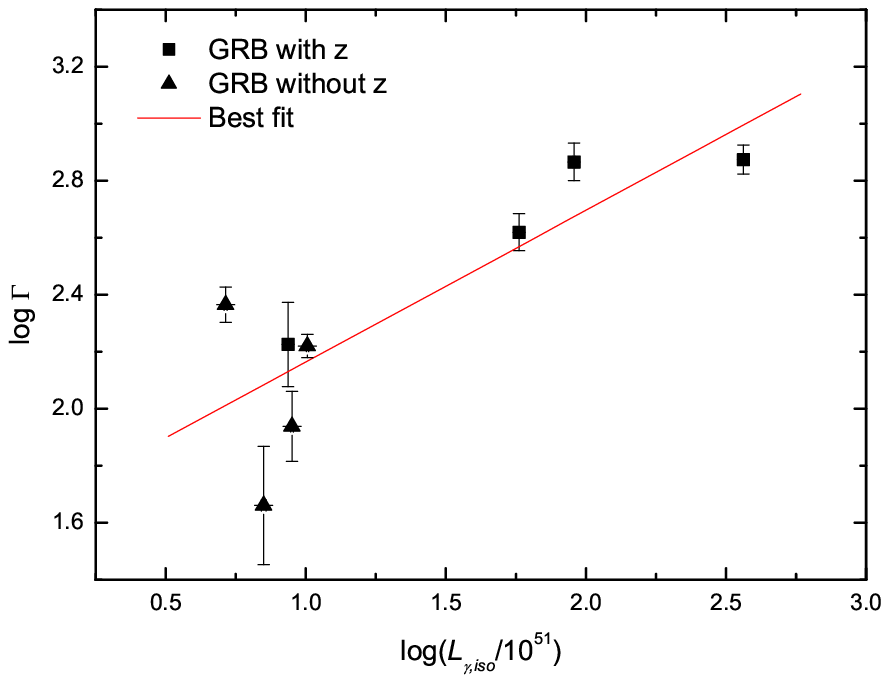}
       \plotone{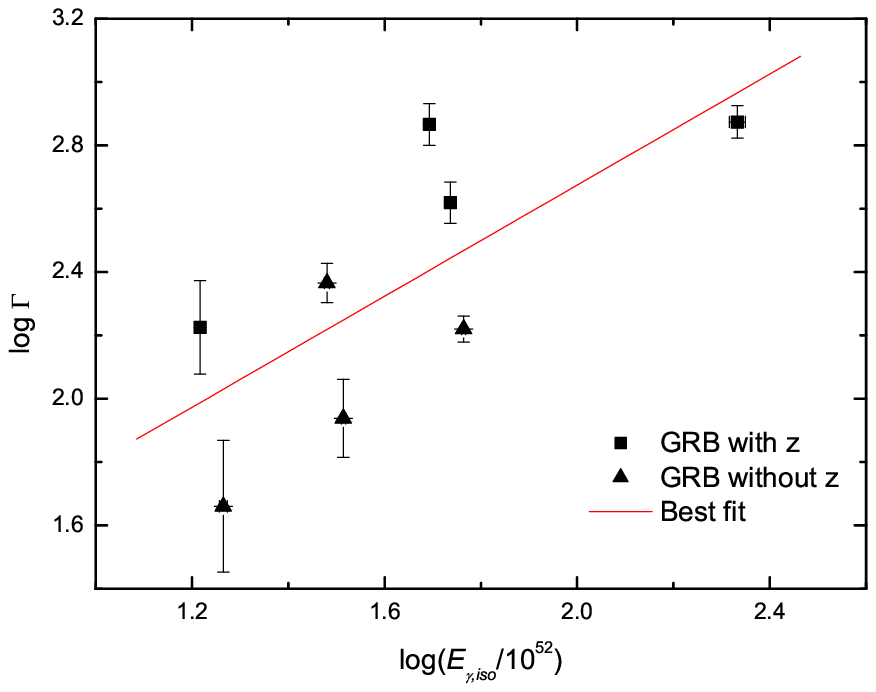}
\hfill \caption{The bulk Lorentz factors as a function of the
isotropic gamma-ray luminosity (top panel) or  isotropic
gamma-ray energy (bottom panel) for 8 bursts with detections of
high-energy spectral cutoffs. GRBs with redshift measurements are
marked with squares and those without redshift measurements
are marked with triangles.} \label{gamma}
\end{figure}

\end{document}